# A Link between Prompt Optical and Prompt Gamma-Ray Emission in Gamma-Ray Bursts

W.T. Vestrand[1], P.R. Wozniak[1], J.A. Wren[1], E. E. Fenimore[1], T. Sakamoto[2], R. R. White[1], D. Casperson[1], H. Davis[1], S. Evans[1], M. Galassi[1], K. E. McGowan[1], J.A. Schier[3], J. W. Asa[3], S. D. Barthelmy[2], J. R. Cummings[2], N. Gehrels[2], D. Hullinger[2], H.A. Krimm[2], C. B. Markwardt[2], K. McLean[1], D. Palmer[1], A. Parsons[2] & J. Tueller[2]

[1] Los Alamos National Laboratory, Space Science and Applications Group, ISR-1, MS-D466, Los Alamos, New Mexico 87545, USA.
[2] NASA Goddard Space Flight Center, Code 661, Greenbelt, MD 20771 USA.
[3] The Pilot Group, 128 W. Walnut Ave., Unit C, Monrovia, CA 91016, USA.

*NOTE: This paper has been accepted for publication in Nature, but is embargoed for discussion in the popular press until formal publication in Nature.*

**The prompt optical emission that arrives with the γ-rays from a cosmic γ-ray burst (GRB) is a signature of the engine powering the burst, the properties of the ultra-relativistic ejecta of the explosion, and the ejecta's interactions with the surroundings[1–5]. Until now, only GRB 990123 had been detected[6] at optical wavelengths during the burst phase. Its prompt optical emission was variable and uncorrelated with the prompt γ-ray emission, suggesting that the optical emission was generated by a reverse shock arising from the ejecta's collision with surrounding material. Here we report prompt optical emission from GRB 041219a. It is variable and correlated with the prompt γ-rays, indicating a common origin for the optical light and the γ-rays. Within the context of the standard fireball model of GRBs, we attribute this new optical component to internal shocks driven into the burst ejecta by variations of the inner engine. The correlated optical emission is a direct probe of the jet isolated from the medium. The timing of the uncorrelated optical emission is strongly dependent on the nature of the medium.**

Starting on 19 December 2004 at 01:42:18 UT, high-energy emission from a bright and very long duration gamma-ray burst, named GRB 041219a, was measured by both the IBIS (**I**mager on **B**oard the **I**NTEGRAL **S**atellite) detector of the INTEGRAL satellite[7] and the Burst Alert Telescope (BAT) of the Swift satellite[8]. The 15-350 keV fluence measured by the Swift BAT was approximately $1.55 \times 10^{-4}$ ergs/cm$^2$, placing it among the top few percent of the 1637 GRB events listed in the comprehensive fourth BATSE (Burst and Transient Source Experiment) catalog[9]. The duration of gamma ray emission from GRB 041219a was approximately 520 seconds, making it one of the longest ever measured.

One of our RAPTOR (**RAP**id **T**elescopes for **O**ptical **R**esponse) telescopes[10] began optical imaging of the GRB 041219a region at 01:44:13 UT, just 8 seconds after receipt of the INTEGRAL alert. The long duration of the burst allowed RAPTOR-S to measure the optical emission in a series of 30-second images for an unprecedented 6.4 minutes while prompt gamma rays were being emitted. At the location of an IR transient identified[11] in subsequent images (starting at 01:49:18 UT) taken by the PAIRITEL telescope, our images show an earlier flash of optical emission (see Figure 1) temporally coincident with the main gamma-ray pulses. At its peak, the optical flash reached a measured magnitude of $R_c$=18.6±0.1 magnitude. However, the location of the event placed it in the galactic plane and in a direction with high optical extinction (galactic longitude and latitude: l=120°, b=+0.1°). Using standard extinction maps[12], we estimate an R-band extinction of ~4.9 magnitudes, but the true extinction may be larger[13]. Correcting for the nominal extinction, the peak flux we measured corresponds to a peak optical magnitude of $R_c$~ 13.7. (Error analysis and our transformation of unfiltered instrumental magnitudes to standard $R_c$-band magnitudes employing standard stars[14] are discussed in a supplementary information file that accompanies this paper on www.nature.com/nature.)

Light curves for prompt optical emission and prompt gamma-ray emission from GRB041219a are shown in the top panel of figure 2. The optical light curve shows: the onset of an optical flash as the dominant first gamma-ray pulse begins, peak brightness during the first gamma-ray pulse, continued optical emission during the secondary gamma-ray peak, and a decay of the optical emission to below our detection threshold during the tertiary gamma-ray enhancement.

Optical emission has been detected during the interval of prompt gamma-ray emission only once before[6], for GRB 990123. Except for an overall temporal scaling factor—GRB 041219a was about 6 times longer—the temporal morphology of the two gamma-ray light curves is remarkably similar (see figure 2). Like GRB 041219a, the gamma ray light-curve for GRB 990123 had a precursor followed by a much larger primary pulse, a secondary pulse, and a smaller amplitude tertiary flux enhancement composed of minor pulses. But in contrast to GRB 041219a, the optical light-curve from GRB 990123 was low during the primary gamma-ray pulse and, though more sparsely sampled, reached peak brightness after the second major pulse. This anti-correlation suggests prompt optical emission from GRB 990123 was generated by a different process than the prompt gamma rays. The consensus interpretation is that the delayed optical peak is generated by a reverse shock [1,3,15], an interpretation supported by detections of the predicted rise to a peak radio flux about one day after the burst [16].

For GRB 041219a, we find the observed optical light curve is well fit by assuming that the generation of prompt optical emission is correlated with the generation of prompt gamma-ray emission. By integrating the observed 15-350 keV flux measured by Swift during the optical exposure intervals and multiplying by a derived constant optical to high-energy flux ratio, we predicted the optical light curve expected if the optical emission and gamma ray emission were perfectly correlated. As shown by the green circles in Figure 3, this simple constant flux ratio assumption predicts both the fast rise of the prompt optical emission observed at the start of the primary gamma-ray pulse and the rapid decline observed after that dominant pulse. Our derived $R_c$-band optical to gamma-ray flux logarithmic color ratio for GRB 041219a is $(R_c-\gamma)= -2.5 \log(F_{opt}/F_\gamma) =17.2$ or, after correcting for an R-band extinction of 4.9 magnitudes, $(R_c-\gamma)=12.3$.

The fast rise of optical emission simultaneous with the dominant gamma-ray pulse, and a general correlation with the prompt gamma-ray emission would naturally arise if emission in both energy bands were generated by a common mechanism. The broadband spectra measured during the optical observation intervals are shown in Figure 4. Modeling of the observed spectra is beyond the scope of this paper, but it can distinguish between emission mechanisms and provide important constraints on physical conditions in the emitting region. A particularly attractive possibility, within the standard internal-external model for GRB fireballs, is that the prompt optical emission observed in GRB 041219a is a low energy tail of the synchrotron emission generated by internal shocks in the GRB outflow[2,30]. In that model, a nearly constant optical to gamma-ray flux ratio requires cooling times short compared to the expansion time, and therefore magnetic fields near equipartition in the ejecta. However, possibilities exist for the emission mechanism, including, for example, saturated Comptonization, which can generate correlated optical and gamma ray emission[17].

Internal shock models[2] typically predict fainter prompt optical emission than reverse shock models. Using the gamma-ray fluxes measured for GRB 990123 and scaling by $1.2 \times 10^{-5}$ (from the (Rc-γ) color derived for GRB 041219a) one predicts significantly lower optical fluxes than measured in GRB 990123—except for the first point in the optical light curve. That first optical measurement, which occurred during the dominant gamma-ray pulse, is consistent, within the prediction uncertainty, with the value predicted for an internal shock using the (Rc-γ) color for GRB 041219a. But after the first measurement, any optical emission generated by internal shocks in GRB 990123 was outshined by bright optical emission from the external reverse shock. To generate the correlated optical and gamma-ray variations measured throughout the full interval of gamma-ray emission in GRB 041219a with internal (forward) shocks, reverse shock emission must be suppressed and/or delayed. The timing and strength of the reverse shock component depends strongly on the physical properties of the relativistic ejecta and the surrounding medium. In fact, the PAIRITEL near IR observations of GRB 041219a show the emergence of a weaker component after the end of the prompt gamma-ray emission that can be interpreted as delayed reverse shock emission[18].

With the addition of the new optical properties displayed by GRB 041219a to the set of known properties for optical emission from GRBs, we can construct a taxonomy of GRB optical emission with three classes: (1) prompt optical emission varying simultaneously with the prompt gamma-rays; (2) early afterglow emission that may start during the prompt gamma-ray emission, but persists for ten minutes or more after the prompt gamma-rays have faded [6,19-21]; and (3) late afterglow emission that can last for many hours to days [22-24]. Within the context of the standard fireball model, it makes sense to attribute the prompt emission to internal shocks in the ultra-relativistic ejecta driven by the GRB engine[2], the early afterglow to a reverse shock driven into the ejecta by interaction with the surrounding medium [1-5,15], and the late afterglow to forward external shocks driven into the surrounding medium generated by interaction with the ejecta [25-26]. This theoretical framework, in turn, allows predictions about the timing, spectra, and relative strength of the optical components that hinge on the properties of the inner engine, the ejecta, and the surrounding medium. The ability of the Swift satellite to provide precise real-time positions and make panchromatic observations of GRBs[27], supplemented by a new generation of sensitive ground-based rapid response telescopes, therefore brings us into a new era in the study of the critical first few minutes during and after GRBs—one that will allow us to probe deeply the physics of these enigmatic explosions.

Acknowledgements: The RAPTOR project is supported by the Laboratory Directed Research and Development program at Los Alamos National Laboratory.

Competing interests statement. The authors declare that they have no competing financial interests.

Correspondence and requests for materials should be addressed to W.T.V. (e-mail: vestrand@lanl.gov).

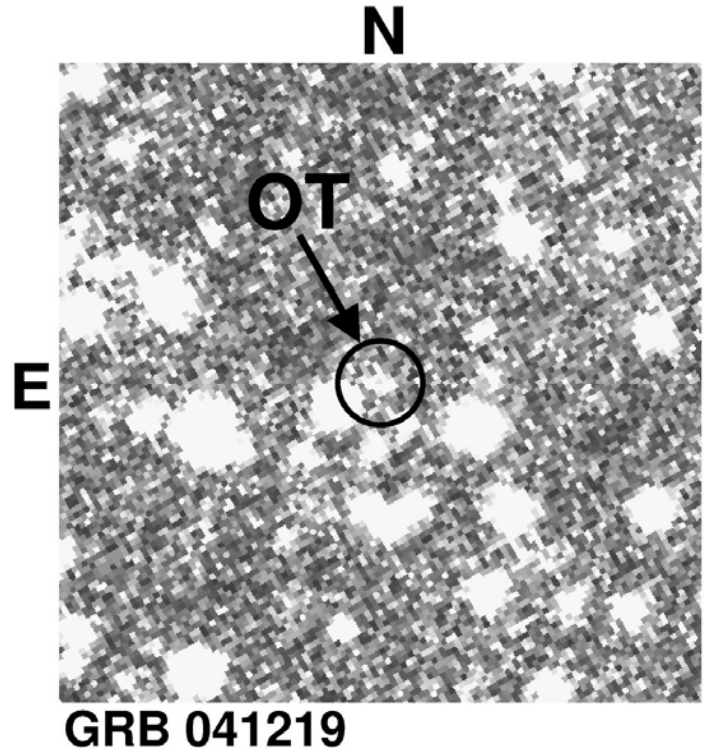

**Figure 1** The prompt optical emission detected from GRB 041219a. This finder chart shows the location (right ascension 00 h 24 min 27.7 s, declination +62° 50′ 33.5″ (J2000)) of the prompt optical flash that we detected, simultaneously with the prompt γ-ray emission detected by the INTEGRAL[7] and Swift[8] satellites, during the time interval 01:45:41–01:49:01 UT on 2004 December 19. The location of the optical transient (OT) is identical to that found both for the subsequent infrared transient[11] and the optical counterpart[28] measured later during the late afterglow phase. Our observations of the prompt optical emission were obtained by RAPTOR-S, a 0.4-m, $f/5$, fully autonomous rapid response telescope owned by Los Alamos National Laboratory and located at an altitude of 2,500 m in the Jemez Mountains of New Mexico. The CCD camera employed for those observations has a 1,056×1,027 pixel format, back-illuminated, Marconi CCD47-10 chip with 13-μm pixels.

| Observing Interval[a] | | $R_c$ Band | 15-25 keV | 25-50 keV | 50-100 keV | 100-350 keV |
|---|---|---|---|---|---|---|
| start | stop | mJy[b] | mJy | mJy | mJy | mJy |
| 202.96 | 275.46 | 2.9 ± 0.8 | 0.74 ± 0.04 | 0.67 ± 0.02 | 0.54 ± 0.02 | 0.28 ± 0.002 |
| 287.95 | 317.95 | 10.3 ± 1.1 | 3.61 ± 0.16 | 2.95 ± 0.09 | 2.18 ± 0.09 | 1.13 ± 0.06 |
| 332.44 | 402.94 | 3.8 ± 0.8 | 2.88 ± 0.07 | 2.02 ± 0.04 | 1.24 ± 0.03 | 0.42 ± 0.03 |
| 415.43 | 573.11 | 1.1 ± 0.5 | 0.58 ± 0.03 | 0.28 ± 0.02 | 0.10 ± 0.02 | 0.0095 ± 0.0072 |

[a] Seconds measured from the event trigger time, $t - t_{trig}$, at 01:42:18.7 UT on 19 December 2004.

[b] Values corrected for 4.9 mag of extinction based on Schlegel et al. maps.

**Table 1** Simultaneous RAPTOR and Swift measurements of GRB 041219a

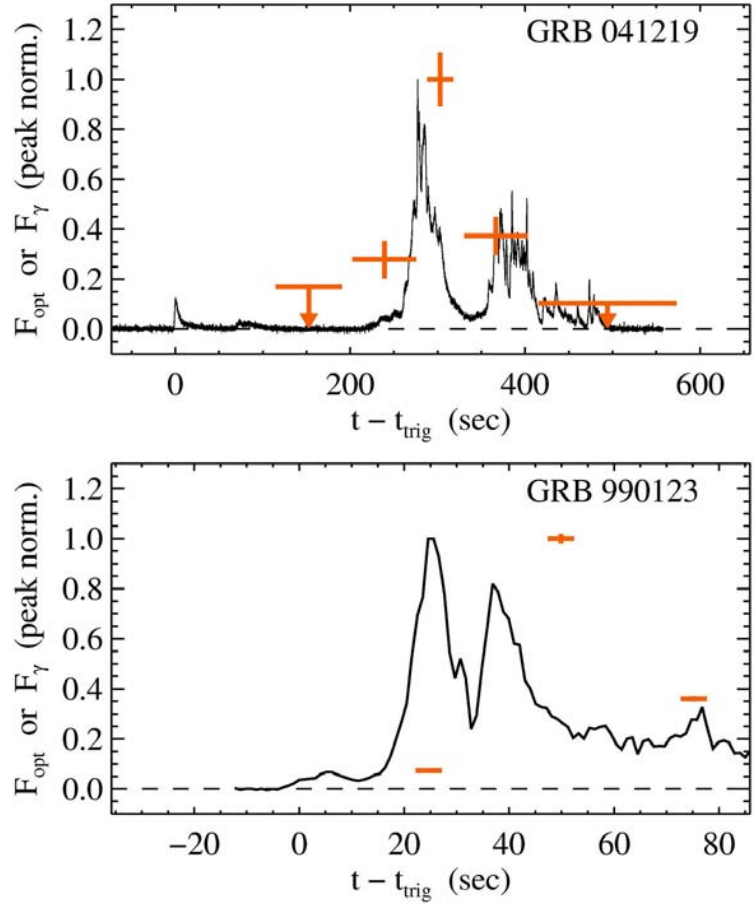

**Figure 2** Comparison of the prompt γ-ray and prompt optical light curves measured[9] for both GRB 041219a and GRB 990123. The black trace in Panel a shows the relative (normalized to the peak value) γ-ray flux, $F_\gamma$, measured for GRB 041219a by the Swift satellite[29] as a function of elapsed time, $t - t_{trig}$, after recognition of the onset of a GRB at trigger time $t_{trig}$=01:42:18 UT. The γ-ray lightcurve for GRB 041219a shows an outburst of precursor emission, followed by a mostly quiet period of 250 s, a primary pulse peaking at about 280 s, a secondary pulse centred at about 380 s, and finally a smaller-amplitude tertiary flux enhancement composed of minor pulses starting at 420 s after the trigger. The black trace in Panel b shows the γ-ray light curve for GRB 990123, which is remarkably similar, except for a temporal scaling factor, to that measured for GRB 041219a. The relative optical fluxes, $F_{opt}$, are indicated by red crosses on both panels with observing intervals denoted by horizontal red lines and $1\sigma$ flux error bars represented by red vertical lines. Together the panels show the relationship between optical emission and γ-ray emission was quite different for the two events. Notice that the early optical flux from GRB 990123 was relatively low during the most intense γ-ray peak and only peaked in the optical[6] after the two primary γ-ray peaks. For GRB 041219a, on the other hand, the prompt optical emission rises rapidly at the start and reaches a maximum during the primary γ-ray pulse, declines but persists during the secondary γ-ray pulse, and then fades below the detection threshold during the tertiary γ-ray enhancement. At peak brightness, GRB 041219a reached $R_c$=18.6±0.1 mag, corresponding to an estimated peak magnitude of $R_c$≈13.7 after correction for extinction by dust.

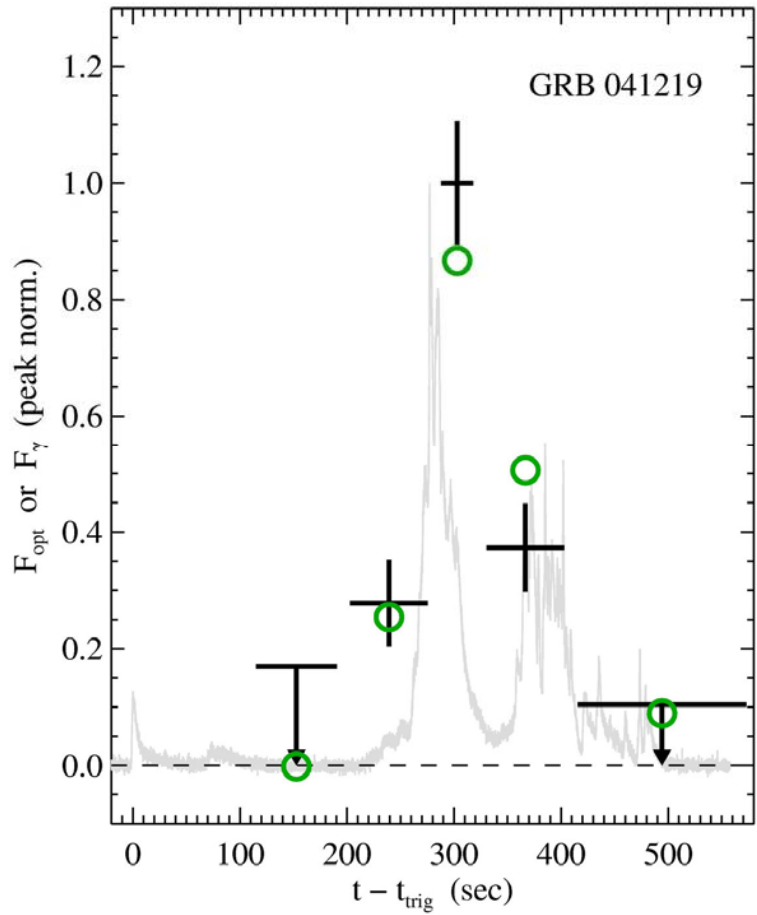

**Figure 3** The measured optical light curve and that predicted for GRB 041219a assuming a constant prompt optical to prompt γ-ray flux ratio. All the optical photometry measurements are derived from stacks of two 30-s images separated by a eight-second readout time, except during the dominant γ-ray peak where a single image yielded a >9$\sigma$ detection. The γ-ray fluxes used for this comparison are derived by integrating the 15–350-keV counting rate measured by the Swift BAT, plotted as the grey trace, during the optical observation intervals. The black crosses show the actual measurements, and the circles show the predicted values. The error bars for detections are given as 1$\sigma$ values, and non-detections are plotted as 2$\sigma$ upper limits. The reduced $\chi^2$ for the best-fitting model for GRB 041219a, with $F_{opt}/F_\gamma$=1.3×10$^{-7}$ (1.2×10$^{-5}$ after correcting for extinction), is $\chi^2$/d.f.=1.79 (4 degrees of freedom, d.f.). In contrast, the best-fitting model employing a constant flux ratio to predict the GRB 990123 optical light curve yields a reduced $\chi^2$ of $\chi^2$/d.f.=1,950.65 (2 d.f.).

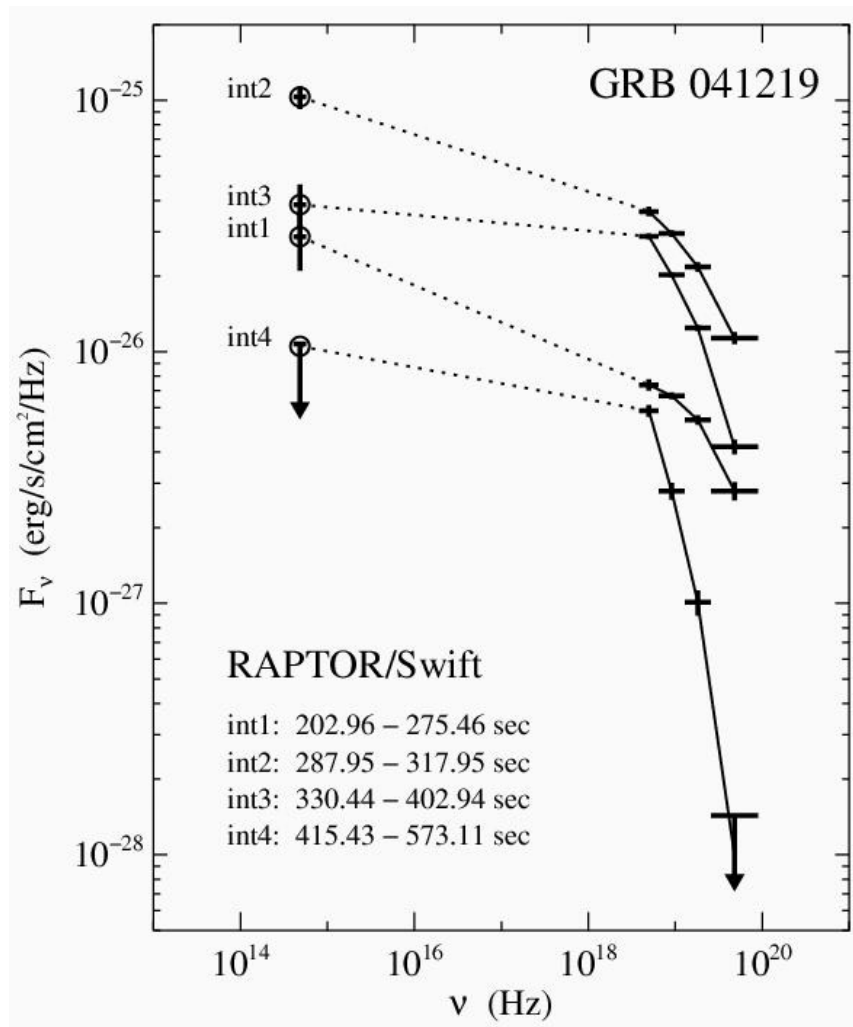

**Figure 4** Broad-band spectra of GRB 041219a, here plotted in flux density $F_\nu$ as a function of observed frequency $\nu$, measured during the period of simultaneous prompt optical and γ-ray emission. The RAPTOR-S optical measurements, after correcting for a nominal 4.9 magnitudes of extinction, are shown as circles. Simultaneous high-energy measurements from the BAT instrument on board the Swift satellite[29] are shown as crosses. All the error bars represent the $1\sigma$ statistical errors. We estimate that the systematic uncertainty for the normalization of the optical fluxes is about a factor of three due to uncertainty in the intrinsic colour of the optical transient and the true extinction along the line of sight. The four integration time intervals, int1–int4, are measured in seconds from the Swift GRB trigger time at 01:42:18.7 UT on 2004 December 19. Notice the optical and γ-ray fluxes vary in concert so that the spectra never cross, and also that the highest-energy band seems to be a slightly better predictor of the behaviour of the optical emission.